\documentclass[prd, twocolumn, lengthcheck, superscriptaddress, showpacs, letterpaper, nofootinbib]{revtex4-1}
\usepackage{amsbsy}
\usepackage{latexsym}
\usepackage{graphicx}
\usepackage{color}
\usepackage{times}
\usepackage{float}
\usepackage{ulem}
\usepackage{amsmath}
\usepackage{hyperref}
\usepackage{microtype}
\usepackage{acronym}
\newcommand\given{\,|\,}

\newcommand\sig{\mathbf{h}}
\newcommand\data{\mathrm{r}}

\newcommand\diff{\, \mathrm{d}}

\def\Msun{\ensuremath{M_{\odot}}}

\def\thercsid{\relax}
\def\rcsid#1{\def\next##1#1{\def\thercsid{##1}}\next}
\rcsid$Id: multiple_pipelines.tex,v 1.51 2012/01/18 17:17:57 jordi Exp $

\renewcommand{\today}{\number\day\space\ifcase\month\or
  January\or February\or March\or April\or May\or June\or
  July\or August\or September\or October\or November\or December\fi
  \space\number\year}

\begin{document}

\title{Detecting transient gravitational waves in non-Gaussian noise with
partially redundant analysis methods}

\author{Rahul Biswas}
\affiliation{University of Texas-Brownsville, Brownsville, Texas 78520, USA}

\author{Patrick R. Brady}
\affiliation{University of Wisconsin--Milwaukee, Milwaukee, WI 53201, USA}

\author{Jordi Burguet-Castell}
\affiliation{Universitat de les Illes Balears, E-07122 Palma de Mallorca, Spain}

\author{Kipp Cannon}
\affiliation{Canadian Institute for Theoretical Astrophysics, University of Toronto, Toronto, Ontario, M5S 3H8, Canada}

\author{Jessica Clayton}
\affiliation{University of Wisconsin--Milwaukee, Milwaukee, WI 53201, USA}

\author{Alexander Dietz}
\affiliation{The  University of Mississippi, University, MS 38677, USA}

\author{Nickolas Fotopoulos}
\affiliation{LIGO - California Institute of Technology, Pasadena, CA 91125, USA}

\author{Lisa M. Goggin}
\affiliation{University of California San Francisco, San Francisco, CA 94143 USA}

\author{Drew Keppel}
\affiliation{Albert-Einstein-Institut, Max-Planck-Institut f\"ur Gravitationsphysik, D-30167 Hannover, Germany}
\affiliation{Leibniz Universit\"at Hannover, D-30167 Hannover, Germany}

\author{Chris Pankow}
\affiliation{University of Wisconsin--Milwaukee, Milwaukee, WI 53201, USA}

\author{Larry R. Price}
\affiliation{LIGO - California Institute of Technology, Pasadena, CA 91125, USA}

\author{Ruslan Vaulin}
\affiliation{LIGO - Massachusetts Institute of Technology, Cambridge, MA 02139, USA}


\begin{abstract}
There is a broad class of astrophysical sources that produce
detectable, transient, gravitational waves.  Some searches for transient
gravitational waves are tailored to known features of these sources.  Other
searches make few assumptions about the sources.  Typically events are
observable with multiple search techniques.  This work describes how to combine
the results of searches that are not independent, treating each search as a
classifier for a given event. This will be shown to improve the overall
sensitivity to gravitational-wave events while directly addressing the problem
of consistent interpretation of multiple trials.
\end{abstract}

\maketitle
\acrodef{LIGO}{Laser Interferometer Gravitational-wave Observatory}

\section{Introduction}
\label{sec:intro}

A variety of astrophysical sources are capable of producing transient
gravitational-wave signals with sufficient strength to be detectable by
ground-based gravitational-wave detectors such as \ac{LIGO}~\cite{Abbott:2007kv}.  Such systems include coalescing
compact binaries (CBC) consisting of neutron stars and/or black
holes~\cite{ratesdoc}. Currently, the same experimental data sets are
searched by several analysis methods. These methods use various signal models and
data processing algorithms and may have different responses to signals and non-Gaussian noise
artifacts~\cite{Slutsky:2009}. Since these searches are not independent, a single, more powerful result can be obtained by combining the results of multiple search methods.

The analysis methods can be divided into two classes by
their assumptions about the signal properties.  The first class assumes that
the waveforms are well modeled and typically employs matched-filtering.  For
this reason, these methods is referred to as template-based searches.  The second class
assumes only basic time frequency properties about the signals.  These methods are
referred to as un-modeled searches. In the template-based searches, there are
often several ways to construct signal models. This means that if a detectable signal exists in the data, it may not perfectly match the signal model chosen for the analysis. For example, the templates may be
constructed using different approximation techniques, or they may correspond to
different parts of the gravitational-wave signal (e.g.~inspiral or ringdown stages of the compact binary coalescence).
The un-modeled searches make minimal assumptions about the shape of
the signal and are designed to detect any short outburst of gravitational
radiation in a given frequency band. Both search classes employ 
algorithms for identifying and discarding the non-Gaussian noise artifacts. To their advantage, un-modeled searches are able to detect a wide class of signals. However, the template-based
searches generally achieve higher detection efficiency for signals matching
the templates.

A gravitational-wave search produces a list of candidate events. Re-analyzing the data with multiple methods may increase the odds of detecting a gravitational wave. At the same time it
has the negative effect of generating redundant lists of gravitational-wave candidates
and increasing the number of trials, which makes it more difficult to
assess the significance of an event. Sensitivity domains of many searches
overlap, meaning that multiple searches may detect the same gravitational-wave signal.
The detection efficiency of a given search depends on a
variety of factors and it can be difficult to
interpret results of multiple searches.

It is apparent that gravitational-wave searches would benefit from
a procedure to consistently combine results into a joint detection
or model exclusion statement for a given population of gravitational-wave sources. 
We apply the general framework for detection of gravitational waves
in the presence of non-Gaussian noise developed in our earlier paper~%
\cite{UWMLSC:2010a} to this problem.  Treating the output of each
search method as a \textit{classifier} for
gravitational-wave candidate events, we construct a unified ranking for all candidate events that is easy to implement and interpret. We test it by combining candidate events from four different search methods that analyze simulated gravitational-wave signals from
compact binary coalescence embedded in the data taken during LIGO's S4
science run. We find that this procedure is robust and can be
used in ongoing and future searches. Interpretation of the combined
results is straightforward. In particular, the calculation of the posterior probability distribution or upper limit for the rate of coalescing binaries can be
carried out as it is normally performed for a single search. The combined upper limit calculation was addressed in \cite{Sutton:2010}, assuming multiple search methods were performed. We briefly discuss the relationship between the method suggested in that paper and ours.

The paper is organized as follows. In Section~\ref{sec:method}, we formulate the
problem of combining results from multiple searches and construct a statistic for the joint analysis. We conclude this section with a discussion of a rate upper limit calculation for the combined search and its relation to the method suggested in \cite{Sutton:2010}. 
In Section~\ref{sec:tests}, we test our procedure by combining results from four different search 
methods. We briefly describe each method, the data, and the model signals.
This is followed by details of the simulations and a discussion of the results. In Appendices
\ref{appendix:JointLR} and \ref{appendix:maxLRapprox}, we derive a formal expression for the multivariate statistic, which accounts for correlations between the searches, and analyze the limits of applicability of our procedure.

\section{Method for combining searches}
\label{sec:method}

In this section, we establish a method for combining results from
different gravitational-wave searches performed over the same data.  The method builds on
the general approach described in \cite{UWMLSC:2010a}.  
We construct a unified statistic for searches by treating each as a separate, 
possibly redundant, classifier for a given event.

Each search method aims to classify observational data into a list of
candidate events, ranked by their likelihood to be a
gravitational-wave signal. In the data analysis process, the data are analyzed
and assigned a rank, $\data$, a real number reflecting the odds that the data contain a gravitational-wave signal. Ordering time series data by amplitude is one simple method for ranking candidate events. The rank (or amplitude) is compared to a pre-established threshold, a boundary that separates signal-like data with sufficient confidence. In
this way, the procedure classifies data on a scale from not signal-like to
signal-like. A search method may classify events by complicated consistency tests and noise rejection schemes, but conceptually any search can be
thought of as a mapping from the space of data to the space of real numbers
that indicate their rank. We will assume that such a ranking procedure
exists for any search method, $i$, and that the result, $\data_i$, indicates
the likelihood that a signal is present in a given search.

Different search methods employ a variety of techniques, data processing algorithms and waveform models. As there are a number of potential gravitational-wave sources, the search targets may vary as well. Separate searches may provide different information about a particular population of sources. Hence, it is important to extract as much information as possible by combining the results of various searches. When multiple searches analyze the same data, the output of each search, $\data_i$, can be further processed to make the most informative detection or rate limit statement for a population of gravitational-wave sources. In doing so, it is important to ensure that there is no loss of detection efficiency when one or more of the methods has a high false alarm rate or is uninformative or irrelevant for the targeted source population. 

For a given event with rank, $\data_i$, one can compute the posterior probability that it is a gravitational-wave signal, $p(1 \given \data_i)$. Following the steps outlined in \cite{UWMLSC:2010a}, this probability can be expressed as
\begin{equation}
\label{probsiggivendata}
p(1 \given \data_i) = \frac{\Lambda(\data_i)}{
\Lambda(\data_i) + p(0) / p(1)} \,,
\end{equation}
\noindent where the likelihood ratio, $\Lambda(\data_i)$, is defined by
\begin{equation}
\label{eq:likelihoodratio}
\Lambda(\data_i) = \frac{\int \! p(\data_i \given \sig, 1) p(\sig \given 1) \diff \sig}
                      {p(\data_i \given 0)} \,,
\end{equation}
\noindent and $p(\data_i \given \sig, 1)$ is the probability of observing $\data_i$ in the presence of the signal $\sig$,
$p(\sig \given 1)$ is the prior probability to receive that signal,
and $p(\data_i \given 0)$ is the probability of observing  $\data_i$ in the
absence of any signal. The targeted astrophysical population of sources is completely described by $p(\sig \given 1)$, where $\sig$ denotes all possible intrinsic (e.g. masses of compact objects in the binary) and extrinsic (e.g. distance to the source, sky location) source parameters. 

If an event is identified as a plausible candidate by several search methods, $p(1 \given \data_i)$ can be calculated for each search based on the ranking, $\data_i$, the event received. Thus, information from each search can be directly compared. The most relevant search results in the highest posterior probability for a signal to be present in the data. According to Eq.~(\ref{probsiggivendata}), this probability is a monotonically increasing function of the likelihood ratio, $\Lambda(\data_i)$. As such, comparing likelihood ratios is equivalent to comparing the posterior probabilities, $p(1 \given \data_i)$. Therefore, the likelihood ratio can be used as a unified ranking statistic to combine the output of all searches.

Strictly speaking, the denominator in  Eq.~(\ref{eq:likelihoodratio}) should contain
contributions from all gravitational-wave sources not included in the targeted population, $p(\sig \given 1)$. We neglect these
terms because typically their contribution is very small. If a
population of sources induces a response very similar to that of the
targeted sources, then these classifiers will not distinguish signals from the
two different populations. This can lead to an overestimation of event rates for the targeted population, however no signal would be missed. Further refinement of
the data analysis techniques or detectors themselves would be required to
distinguish between the signals from these sources.

We define the ranking statistic for the joint search to be
\begin{equation}
\label{maxLRapprox}
\data_\mathrm{joint} = \max
\left\{\Lambda(\data_1),\Lambda(\data_2),\ldots,\Lambda(\data_n)\right\}\,,
\end{equation}
\noindent where maximization is carried out over simultaneous events. Though this choice does not make use of all available information (we neglect correlations between the classifier's ranks; see Appendix \ref{appendix:JointLR} for multivariate treatment of the problem), it does
offer some advantages. It is straightforward to
compute $\Lambda(\data_i)$ for each search
method and simply take the largest.  This has a simple
interpretation: events from each search are compared based on the ratio of sensitivity
of the method to the targeted sources and the
search's false alarm rate. The event that is most likely to be a
gravitational wave is kept. As a result, the searches are combined according to the best classifier for each
event. Events classified by noisy,
insensitive searches receive a low
likelihood-ratio ranking and therefore do not contaminate the overall
sensitivity of the analysis. We further discuss the limits of applicability for this ranking statistic in Appendix \ref{appendix:maxLRapprox}. 

As in the case of a single search method, the result
of combining searches using the maximum likelihood-ratio statistic,
Eq.~(\ref{maxLRapprox}), is a list of events.  They
can be treated as the output of a single search, with their 
significance evaluated by estimating the background. In the next
section, we discuss how to do this. Having a single list of
events allows for straightforward interpretation of
results.  The most significant events can be further
studied and possibly promoted to the list of detected gravitational-wave
signals. The posterior probability distribution or the upper limit on the rate
of coalescence can be calculated following any of the methods developed for a single
search \cite{Feldman:1997qc, loudestGWDAW03,
Biswas:2007ni}.

The upper limit calculation for multiple searches described in \cite{Sutton:2010} differs from our method. In \cite{Sutton:2010}, searches are treated as counting
experiments and the upper limit on the rate of events is calculated using the
total number of events  above some fixed threshold
and Poisson statistics. To apply this method when multiple searches are performed, a prescription is needed for determining
how many events each search should contribute to the total count. In \cite{Sutton:2010} events are classified by combinations of searches that generated them. The problem is reduced to the choice of foliation by
a family of exclusion surfaces, $S(\zeta)$, of the space of the number of events in each category, $N_q$, where $q$ runs through all possible
combinations of $m$ $( m \leq n)$ out of $n$ searches. In the paper, the authors suggest and discuss several
plausible choices of linear surfaces, $S(\zeta)$, that lead to different upper
limits.
\noindent By construction, the maximum likelihood ratio, Eq.~(\ref{maxLRapprox}), ensures that the total number of
events each search contributes on average to the joint search is proportional to the ratio of its efficiency to
detect the targeted signals to its background. This construction is closely
related to the ``single combination'' option of \cite{Sutton:2010}, in which only the most sensitive search contributes to the upper limit. Note though, that in our method the most sensitive search is determined during the analysis on event by event basis. This relieves an analyst from determining before hand which of the searching methods is the most sensitive. Often sensitivity is a very complicated function of signal's parameters and there might not be a single most sensitive search method that covers all signals. In this case, one would have to split the signal parameter space into subdomains, within which a single most sensitive search method exist, and carry out the upper limit calculation for each of the domains independently. In practice, this may prove to be a formidable task. The maximum-likelihood ratio procedure is universal and is almost trivial to implement, as we show it the next section.  Its other important advantage is accounting for background noise present in each of the search methods. The choice between the methods is based not only on their sensitivity but also their susceptibility to the noise artifacts. In \cite{Sutton:2010}, the authors also mention the necessity to include
information about the background to achieve more optimal upper limits.  

The maximum likelihood construction ignores correlation between the searches. The optimal way to account for it is to define the multivariate likelihood-ratio ranking described in Appendix \ref{appendix:JointLR}. Unfortunately, implementing this ranking for more that two search methods is not feasible. Also, we argue that in most practical situations the net positive effect of correlations is small. The multivariate likelihood-ratio, Eq.~(\ref{LRGenFormula}), defines the optimal exclusion surfaces, $S(\zeta)$, for the upper limit calculation method of \cite{Sutton:2010}. These surfaces generally are non-linear and, therefore, do not directly correspond to any of the choices considered in \cite{Sutton:2010}. The closest in spirit is the ``efficiency-weighted combination'' suggested by the authors of \cite{Sutton:2010}, where contributions from each combination of the search methods are weighted proportionally to their sensitivity to signals.  Using notation of \cite{Sutton:2010} and accounting for noise contribution, the corresponding exclusion planes are defined by the normal vector $\mathbf{k} \equiv
(\epsilon_q/b_q)$,  where $\epsilon_q$ is the probability of a signal to be
detected by the $q^\mathrm{th}$ combination of the searches
and $b_q$ is the number of background events in
this combination.

\section{Testing the maximum likelihood-ratio statistic with non-Gaussian data}
\label{sec:tests}

The maximum likelihood-ratio statistic, Eq.~(\ref{maxLRapprox}), provides a natural way to combine results of several search methods into a single joint search. It possesses several attractive qualities and is expected to result in no loss of efficiency in the most practical situations (see Appendix \ref{appendix:maxLRapprox} for discussion of this). Still, it is important to verify this in conditions that mimic the strong non-Gaussian noise that is encountered in the search for gravitational waves in real data. To simulate a real life application of our procedure, we employ four search methods that are currently used to detect gravitational waves from coalescing binaries with the LIGO and Virgo observatories. We analyze simulated signals inserted in the data from LIGO's fourth scientific run (S4) and combine results of these analyses using the maximum likelihood-ratio statistic. We estimate the efficiency of the combined search in detecting these signals in the typical LIGO noise and compare it to the efficiencies of the individual searches.

\subsection{Data and signals}

We insert simulated signals into data collected between February 24 and March 24, 2005, during LIGO's S4 run. The data was taken by three detectors: the H1 and H2
co-located detectors in Hanford, WA and L1 in Livingston, LA\@. Several
searches for gravitational waves were performed in these data, but no
gravitational-wave candidates were identified \cite{Abbott:2009km, LIGOS4burst,
LIGOS3S4all}. For this work, 15 days of triple coincidence data were used, which is the sum of all times during S4 when all three detectors were simultaneously operating in science mode.

Our signals include three kinds of binaries: neutron star--neutron star (BNS), neutron
star--black hole (NSBH) and black hole--black hole (BBH) binaries. We use
non-spinning waveforms to model signals from these binaries. For BNS,
these are post-Newtonian waveforms~\cite{Blanchet:1996pi,Droz:1999qx,Blanchet:2002av,Buonanno:2006ui,Boyle:2007ft,Hannam:2007ik,pan:024014,Boyle:2009dg,thorne.k:1987,SathyaDhurandhar:1991,Owen:1998dk},
Newtonian order in amplitude and second order in phase, calculated
using the stationary phase approximation~\cite{Droz:1999qx,
thorne.k:1987,SathyaDhurandhar:1991} with the upper cut-off frequency set by the
Schwarzschild innermost stable circular orbit (ISCO). Signals from all other
binaries are approximated by the effective one body numerical relativity
(EOBNR)
waveforms~\cite{BuonannoDamour:1999,BuonannoDamour:2000,DamourJaranowskiSchaefer:2000,Damour03,Buonanno:2006ui,Pan2007,Buonanno:2007pf,DN2007b,DN2008,Boyle2008a,Damour2009a,Buonanno:2009qa}.
The former waveforms describe only the inspiral phase of the coalescence,
whereas the latter also include merger and ringdown phases.

The simulated signals are injected into non-overlapping 2048-second
blocks of data. To improve the statistic, multiple signal populations are
inserted in the data and independently analyzed. Signals are split into three
categories by total mass of the binary: 2--6~\Msun, 6--100~\Msun, and
100--350~\Msun. The lowest mass range includes only BNS systems. Within each mass
range, signals are distributed uniformly in distance or the inverse of distance.
The BNS range covers 1--20~Mpc while other systems reach from 1--200~Mpc. In order to represent realistic astrophysical population with probability density function scaling as distance squared, the simulated signals are appropriately re-weighted and are counted according to their weights. All other
parameters of the signals have uniform distribution. In total, there are 943,
2245, and 2237 signals injected within each mass category, respectively.

\subsection{Search methods}

Four search methods, each representing one of the standard searches for transient gravitational-wave signals in LIGO and Virgo data, are used to perform this joint analysis. Brief descriptions of the search methods are given below. The first three are template-based searches, while the last one does not rely on any specific signal model.

The low-mass CBC pipeline targets binaries with total mass below 35~\Msun. The data from each interferometer are match-filtered with a bank of non-spinning post-Newtonian waveforms~\cite{Blanchet:1996pi,Droz:1999qx,Blanchet:2002av,Buonanno:2006ui,Boyle:2007ft,Hannam:2007ik,pan:024014,Boyle:2009dg,thorne.k:1987,SathyaDhurandhar:1991,Owen:1998dk} covering binary mass combinations with total mass in the range 2--35~\Msun. The template waveforms are calculated in the frequency domain using the stationary phase approximation~\cite{Droz:1999qx,thorne.k:1987,SathyaDhurandhar:1991} to Newtonian order in amplitude and second PN order in phase. The waveforms are extended up to the Schwarzschild ISCO\@. When the signal-to-noise ratio (SNR) time series for a particular template crosses the threshold of 5.5, a single-interferometer trigger is recorded. These triggers are required to pass waveform consistency and coincidence tests with triggers from other interferometers. The surviving triggers are ordered by a ranking statistic and form a ranked set of candidate events. For detailed descriptions of this pipeline and recent search results, see~\cite{LIGOS3S4all, LIGOS3S4Tuning, Collaboration:2009tt, Abbott:2009qj, S5LowMassLV}.

The high-mass CBC pipeline is similar to its low-mass counterpart, however it is designed to target binaries with total mass between 25--100~\Msun. The EOBNR family of templates used in a high-mass search has waveforms covering the evolution of a coalescing binary from late inspiral to ringdown. Other than the choice of templates, the analysis is quite similar to the low-mass search. The high-mass CBC pipeline was used to search for gravitational waves from binary black holes in the S5 LIGO data~\cite{Collaboration:2010a}.

The ringdown pipeline was developed to search for gravitational-wave signals corresponding to the post-merger phase of the binary coalescence. After two compact objects merge, a single, highly perturbed black hole forms and radiates energy while it settles down to a stable Kerr solution. The pipeline constructs its template bank from the dominant, $l=2$ and $m=2$, black hole quasi-normal modes characterized by a single frequency and quality factor. The template bank used in the ringdown search spans the most sensitive part of the LIGO frequency band, 50 Hz--2 kHz. Quality factors between 2--20 are used, corresponding to a range for the final black hole spin between non-spinning and ${\hat a} = 0.994$. As with the other search methods previously described, candidate events are ranked after being detected by multiple interferometers and passing several consistency tests. This pipeline was used to search for gravitational waves in the S4 data~\cite{Abbott:2009km}.

Coherent WaveBurst is a gravitational-wave burst analysis pipeline designed to detect signals from transient gravitational-wave sources. It uses minimal information about the signal model, instead using the cross-correlated excess power from the gravitational-wave signal across a network of interferometers~\cite{Klimenko:2007hd}. The pipeline enforces the signal hypothesis by maximizing a likelihood functional that describes the expected signal response of an impinging gravitational wave given its source location in the sky combined over the network of interferometers~\cite{Klimenko:2005wa}. Triggers are generated from the interferometer network by combining time-frequency maps using wavelet transformations of the interferometer time-series data. From these maps, the likelihood of a trigger is calculated by the pipeline from the correlation of the whitened data streams weighted by the network's antenna patterns. This pipeline was used in the LIGO S4~\cite{Abbott:2008eh} and LIGO/Virgo S5/VSR1~\cite{abbott-2009b,S5VSR1Burst} searches for un-modeled short duration transients, as well as searches for black hole binaries~\cite{Pankow:2009lv}.

We note that our analysis does not include the most recent innovations developed to improve the efficiency of each of these pipelines. In particular, we do not categorize the candidate events by the template mass and coincidence type -- a novelty introduced in the low- and high-mass CBC pipelines during the analyses of S5 LIGO data and the S5/VSR1 data from LIGO and Virgo. Also, we use the default settings for the S5 analysis (S4 for the ringdown search) for numerous pipelines' parameters without attempting to re-tune them. We choose to perform simulations without the most up-to-date and fully tuned versions of the pipelines to save time. This is justified because our algorithm for combining searches is ignorant of the inner-workings of each pipeline. This makes the combined search robust against small changes in the individual analysis algorithms. For our purpose, it is sufficient to use somewhat simplified versions of the pipelines, as we do not expect these results to change dramatically when incremental changes occur as the pipelines evolve.

\subsection{Algorithm for combining searches}\label{sec:tests_algorithm}

The procedure for combining candidate events identified by different classifiers is straightforward and based on Eq.~(\ref{maxLRapprox}). The first step is the calculation of the likelihood ratio,
$\Lambda(\data_i)$, defined by Eq.~(\ref{eq:likelihoodratio}), for every event.
Notice that since the numerator depends on
the population of signals through $p(\sig \given 1)$, $\Lambda(\data_i)$ is not just a trivial re-scaling of
a rank assigned by a classifier to an event. The likelihood ratio estimates the significance of each event in the context of a gravitational-wave detection from the targeted population of sources specified
via  $p(\sig \given 1)$ and $p(\data_i \given \sig, 1)$.  As a result, events are ranked by the odds of being produced by the classifier in response
to the targeted signals, rather than noise. Depending on the population of
sources, some classifiers may not provide any useful information.
In that case, events provided by such classifiers receive a very low likelihood
ratio rank and are effectively removed from the search. This is a desirable
feature that makes the procedure robust against nuisance classifiers.

In order to compute the one-dimensional likelihood ratio given in Eq.~(\ref{eq:likelihoodratio}), we
need to measure the classifiers' response to the gravitational-wave signals
interposed over noise and to background noise only. For the latter, we use a common background estimation technique for
gravitational-wave searches --- shifting recorded  data from the non-colocated interferometers in time with
respect to each other~\cite{LIGOS3S4all, LIGOS3S4Tuning, Collaboration:2009tt,
Abbott:2009qj, S5LowMassLV}. If the shift is much longer than the
gravitational-wave travel time between the detector sites ($\approx 10$ ms for a
Hanford-Livingston detector pair),
then the resulting time-shifted data are guaranteed to contain no coherent
gravitational-wave signals. These data, when analyzed by the classifier, represent
the background of the search. In the low-mass CBC, high-mass CBC, and ringdown pipelines, we perform 2000 forward-in-time shifts of the L1 data with time steps of 7 seconds relative to H1 and H2. The Coherent WaveBurst search uses 100 forward-in-time shifts of the L1 data with 5-second time steps. Each time-shifted data set produces an
independent sample of the background. The time-shifted data are
analyzed and all background events are recorded.

For background events, it is significantly easier to estimate the cumulative probability density function (cdf), $P(\data_i \given 0) = \int_{\data_{i}}^\infty \! p(\data'_i \given 0)\diff \data'_i$, than the probability density function (pdf), $p(\data_i \given 0)$. Each background data set represents an independent trial observation of duration, $T$. Therefore, for a trigger, $\data_i$, the ratio of the number of the trial observations that produced a trigger with the rank $\data'_i \geq \data_i$ to the total number of trial observations provides an estimate of the probability, $P_{T}(\data_i \given 0)$, of the classifier producing an equally or higher ranked event in the analysis of noise alone. This probability is a monotonic function of $P(\data_i \given 0)$ and experiment duration, $T$,
\begin{equation}
P_{T}(\data_i \given 0) = 1 - (1 - P(\data_i \given 0))^{T/T_0} \,,
\end{equation}
where $T_0$ is the duration of a unit experiment, which can be classified by a single rank, $\data_i$. The scale for $T_0$ is set by the duration of the gravitational-wave signal, the time scale on which data samples can be considered uncorrelated. In practice, given that all methods analyze the same amount of data, the two probabilities $P_{T}(\data_i \given 0)$ and $P(\data_i \given 0)$ are equivalent for the purpose of ranking the candidate events since one is a monotonic function of the other. Computation of the background cdf curve, $P_{T}(\data_i \given 0)$, is a trivial task. First, the single event with the highest rank from every background data set is chosen. Then, for any value of $\data_i$, one simply counts the number of these events with rank $\data'_i \geq \data_i$, divided by the total number of background data sets. In this way, the background cdf curves are calculated for each search method.

In order to measure the response of each search method to the targeted gravitational-wave signals and calculate the numerator in Eq.~(\ref{eq:likelihoodratio}), several populations of simulated
signals are injected into the data and processed by the pipelines. For
each search method, events identified with the injected signals are recorded
along with the parameters of the signals. As with background events, it is
much easier to compute the cdf, $P(\data_i \given 1)$, of the signals. For an event with rank $\data_i$, this probability is approximated by the ratio of the number of injected signals with rank
$\data'_i \geq \data_i$ to the total number of injected signals. Using this
algorithm we compute cdf curves, $P(\data_i \given S_j, 1)$,  for each search method,
$i$, and mass category of the simulated signals, $S_j$: 2--6~\Msun, 6--100~%
\Msun, and 100--350~\Msun, which represent the intended targets of the low-mass CBC, high-mass CBC, and burst/ringdown pipelines, respectively.

Having pre-computed the background and signal cdf curves, the algorithm for
combining the analysis pipelines can be summarized in the following two 
steps. First, every event, $\data_i$, produced by the $i^\mathrm{th}$ classifier is
assigned the log-likelihood-ratio ranking given by
\begin{equation}
\label{rankingLR}
L(\data_i \given S_j) = \ln\left[\frac{P(\data_i \given S_j, 1)}{P_{T}(\data_i \given 0)}\right] \,,
\end{equation}
in which the ratio of pdfs  in Eq.~(\ref{eq:likelihoodratio}) is approximated by the ratio of
cdfs. For the sake of brevity in what follows we omit the ``log'' from the ranking name and refer to $L(\data_i \given S_j)$ as the likelihood-ratio ranking. Second, all events  from all classifiers are mixed together and clustered,
retaining events with the highest likelihood-ratio ranking, $L(\data_i \given S_j)$,
within the specified time window. These events form the final list of
gravitational-wave candidates. The time window is approximately equal to
the autocorrelation time for an average signal injected in the data. In our
simulations, we set it to 10 seconds. Events separated in time by more than 10
seconds are uncorrelated and therefore may correspond to different signals~\cite{LIGOS3S4all, LIGOS3S4Tuning, Collaboration:2009tt, Abbott:2009qj, S5LowMassLV}.
This last step effectively implements maximization in Eq.~(\ref{maxLRapprox})
over the likelihood ratio for coincident events identified by multiple search methods.

We expect the cdf approximation used in Eq.~(\ref{rankingLR}) to be fair in the context
of our simulations. Injection and background distributions are one-dimensional,
monotonic functions of rank. They generally fall off as some
negative power of rank. Detectable signals lie on the tail
of the background distribution. Under these conditions, the difference between using the pdf or
cdf in the likelihood ratio is insignificant. Nevertheless, we should stress
that this may not be the case in general and proper estimation of signal and
background probability distributions may be required.

Before proceeding to the discussion of simulation results, we note
that $L(\data_i \given S_j)$, defined by Eq.~(\ref{rankingLR}), depends on the population of injected signals, $S_j$.
Therefore, events identified by the classifiers in each search must be re-processed according to the algorithm described above for each population of sources, $S_j$. 

\subsection{Simulation results}

After multiple search methods are used on the data injected with gravitational-wave signals, the events selected by each search are processed with the algorithm sketched in the previous subsection. To estimate the background for the combined search, we again perform time shifts of the L1 data with respect to data from H1 and H2. Although the time shifts performed in \ref{sec:tests_algorithm} are independent for each classifier, the time shifts must be synchronized for all classifiers when estimating the background for the combined search. For this purpose, 100 5-second time shifts of the L1 data are performed with respect to the H1 and H2 data. The background sample is processed with the same algorithm as the main data.

As previously mentioned, we consider three target populations of
compact binaries, categorized by their total mass: the binary neutron stars with
total mass 2--6~\Msun, the compact binaries with total mass 6--100~\Msun, and
the binaries with total mass 100--350~\Msun. These define three independent
searches. The data with injected signals from each category are analyzed independently. The resulting events are ranked by the
likelihood ranking, Eq.~(\ref{rankingLR}), with $S_j$ being one of the considered
target populations. The background events are ranked and combined in the
same way, providing an estimate of the background for the combined searches.

\begin{figure}
\includegraphics[width=3in]{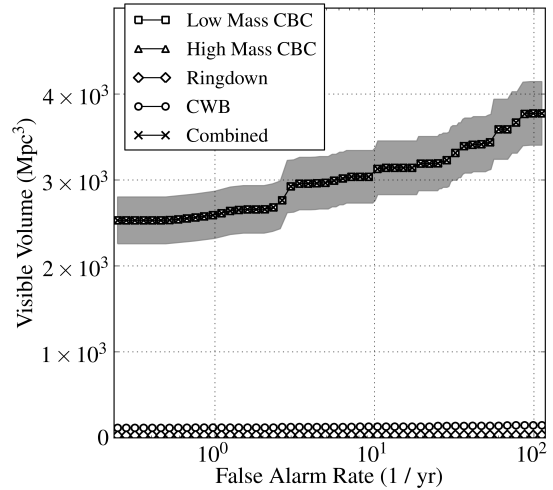}
\caption{Visible volume versus false alarm rate for binary neutron stars. The shaded area around a curve represents its 1$\sigma$ Poisson error. The ``Combined'' and the ``Low Mass CBC'' curves coincide, whereas all other curves drop to near zero in visible volume.}
\label{fig:bns_LVFAR}
\end{figure}

\begin{figure}
\includegraphics[width=3in]{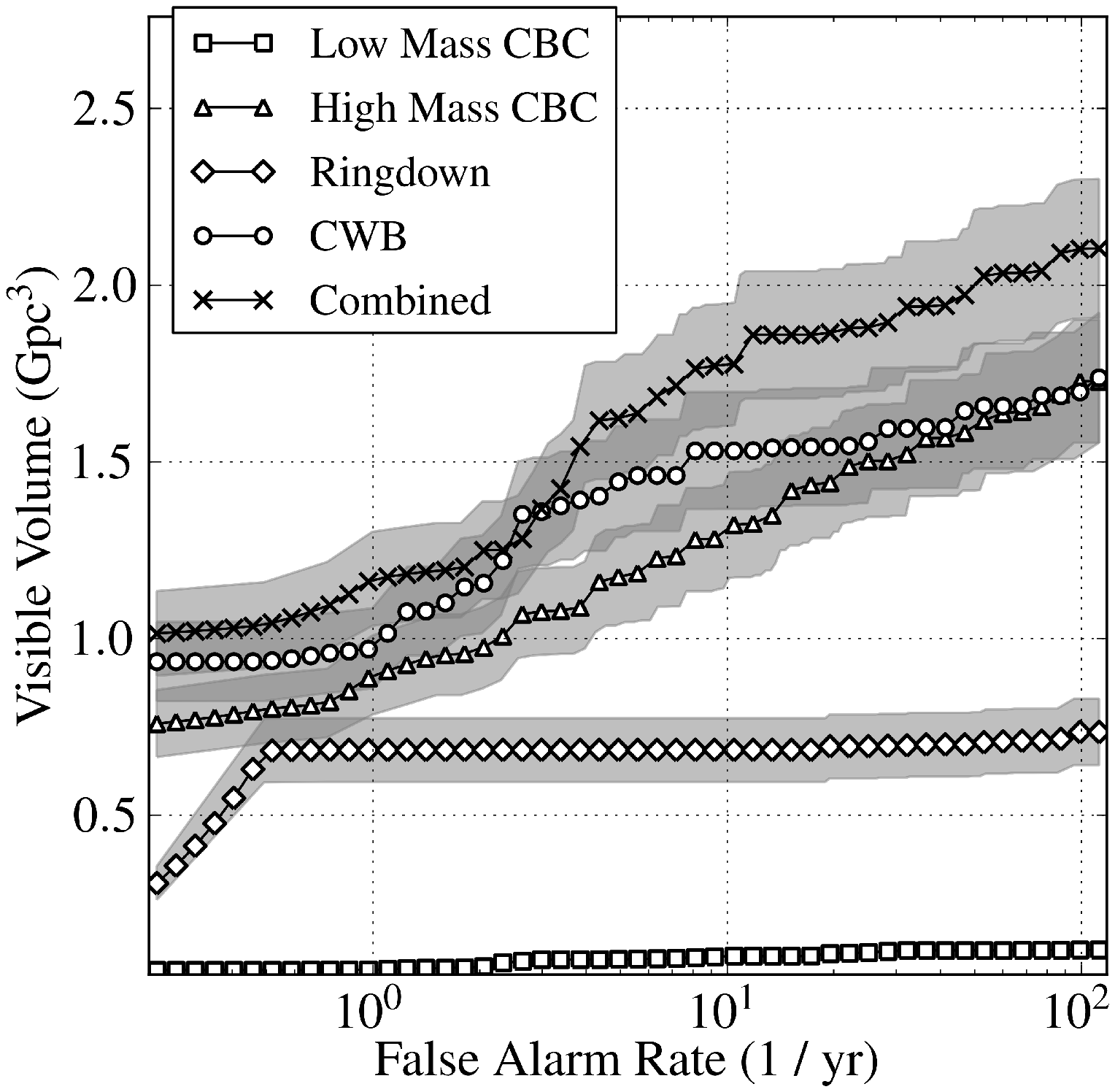}
\caption{Visible volume versus false alarm rate for CBC with total mass 6--100~\Msun. The shaded area around a curve represents its 1$\sigma$ Poisson error.}
\label{fig:medmass_LVFAR}
\end{figure}

\begin{figure}
\includegraphics[width=3in]{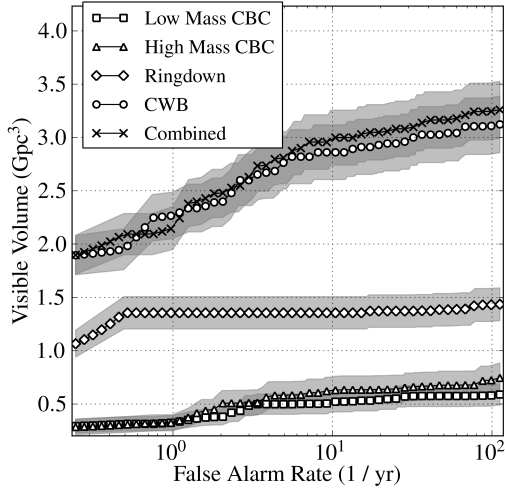}
\caption{Visible volume versus false alarm rate for CBC with total mass 100--350~\Msun. The shaded area around a curve represents its 1$\sigma$ Poisson error. }
\label{fig:highmass_LVFAR}
\end{figure}

To compare combined searches with the individual search methods, we compute their
sensitivities to the signals in the presence of typical background. We
summarize this in Figures \ref{fig:bns_LVFAR}--\ref{fig:highmass_LVFAR}, which
show curves of visible volume versus false alarm rate for each of the search methods and for the combined search. For each point on these curves, the calculation proceeds as follows. First, using background events, we determine the value of the rank corresponding to a given false alarm rate. Next, the efficiency as a function of distance to the source, $\epsilon(D)$, is estimated by the fraction of the signals at distance $D$ ranked above that value. This efficiency is then converted to the visible volume.
\graphicspath{{multiple_pipelines_plots/}}

The binary neutron star search is a case study in which only one of the classifiers, namely the low-mass CBC pipeline, is effective in detecting the particular type
of gravitational-wave signal. All other classifiers are
designed to detect either black hole binaries or short duration bursts and
are inefficient in detecting the long inspiral signal sweeping through the whole
LIGO frequency band. This is properly accounted for in the likelihood-ratio ranking, which
is very low for all events from these other classifiers. The only events
not de-weighted are those from the low-mass CBC pipeline. As a result, the
combined search is equivalent to the low-mass CBC search in this case. The sensitivity curves for both searches, shown on Figure~\ref{fig:bns_LVFAR}, coincide. This shows that our algorithm
is robust against uninformative, nuisance classifiers.

The picture changes dramatically for compact binaries in the medium mass range, shown in 
Figure~\ref{fig:medmass_LVFAR}. In this case, the efficiency of the low-mass CBC
pipeline is negligible in comparison to the other classifiers. In this category, the Coherent WaveBurst
pipeline has the best overall sensitivity. Further inspection reveals that the
high-mass CBC pipeline is the most sensitive of the three in the 6--50~\Msun{} mass
region, whereas the ringdown pipeline, despite being subdominant, tends to
detect signals with high mass ratio that are either missed or not ranked
high enough by the other pipelines. Thus, in this case, all but one classifier contribute
detected signals to the combined  search (the ``$\times$--$\times$'' curve  on
Figure~\ref{fig:medmass_LVFAR}). This is a desired effect of incorporating the
detection sensitivities of different pipelines, which results in a more
sensitive and robust combined search.

We observe similar effects for the high-mass binaries,
although this is not obvious from Figure~\ref{fig:highmass_LVFAR}. The figure shows
the Coherent WaveBurst pipeline dominating over the
ringdown or the high-mass CBC pipelines. The sensitivity curve of the combined
search tends to be just above the Coherent WaveBurst curve and occasionally drops below it. However, these drops are well within the error bars. The detailed investigation shows that the
high-mass CBC and the ringdown pipelines still contribute detections of extra
signals missed by the Coherent WaveBurst pipeline. In particular, the ringdown pipeline has the
highest sensitivity of all searches in the 270--350~\Msun{} region. Most of these
extra signals are in the near or mid range zone (less then 100~Mpc) and
therefore do not contribute as much to the total visible volume as those at far
distances. As a result, overall gain for the combined search is not that
significant when compared to the Coherent WaveBurst search alone. Moreover, occasionally, due to
background fluctuations, the threshold for the combined search fluctuates upward, which
results in a loss of a few distant signals detected by the Coherent WaveBurst pipeline. The loss of visible volume associated with these signals
is not compensated by the gain of efficiency in the mid range. We should note that the measurement of detection efficiency for signals beyond 150 Mpc has
large uncertainties due to low counts for detected signals. Therefore, the actual
loss of efficiency in this case may be overestimated.

\begin{figure}[ht]
\includegraphics[width=3in]{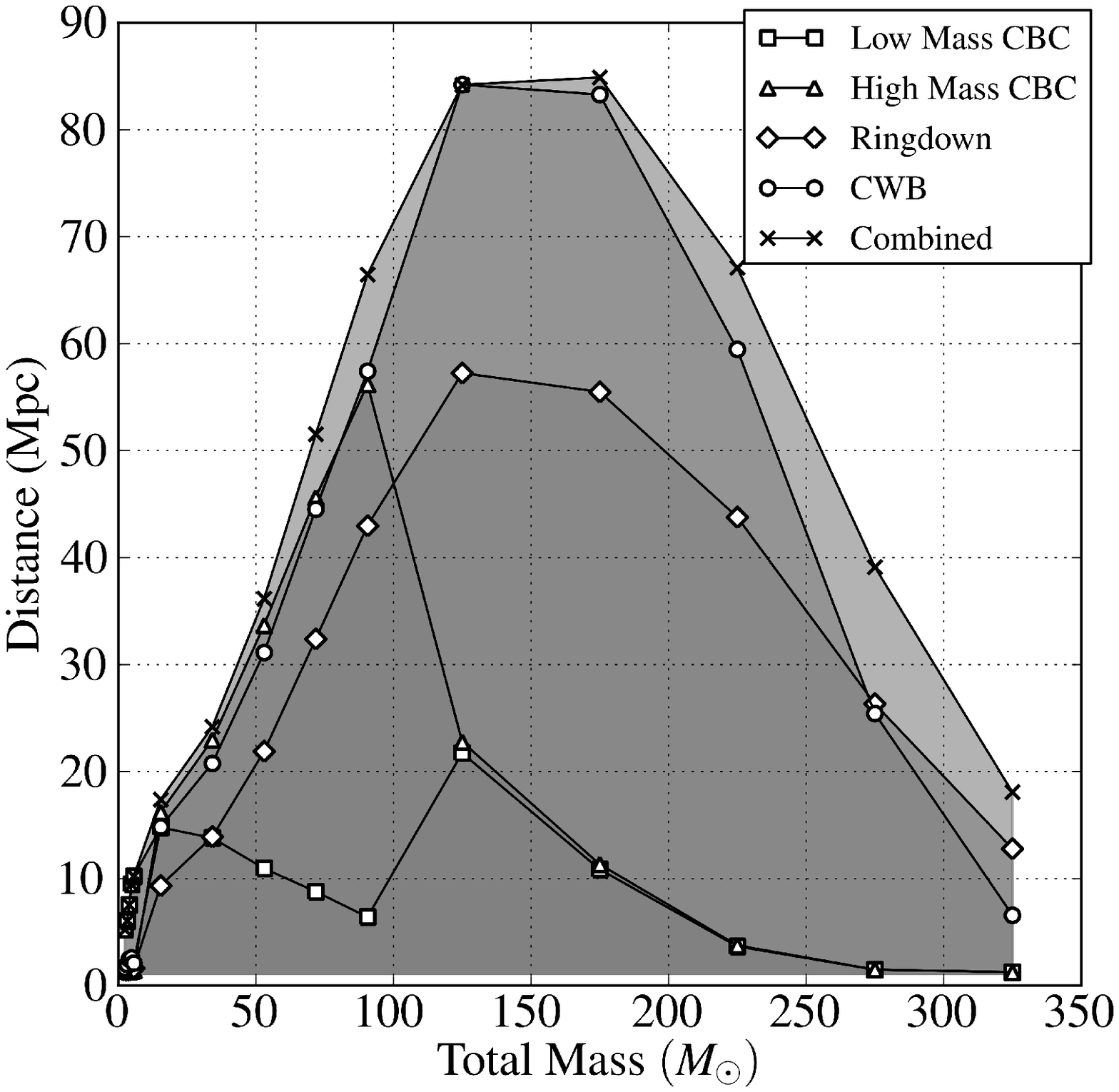}\vspace*{0.15cm}
\includegraphics[width=3in]{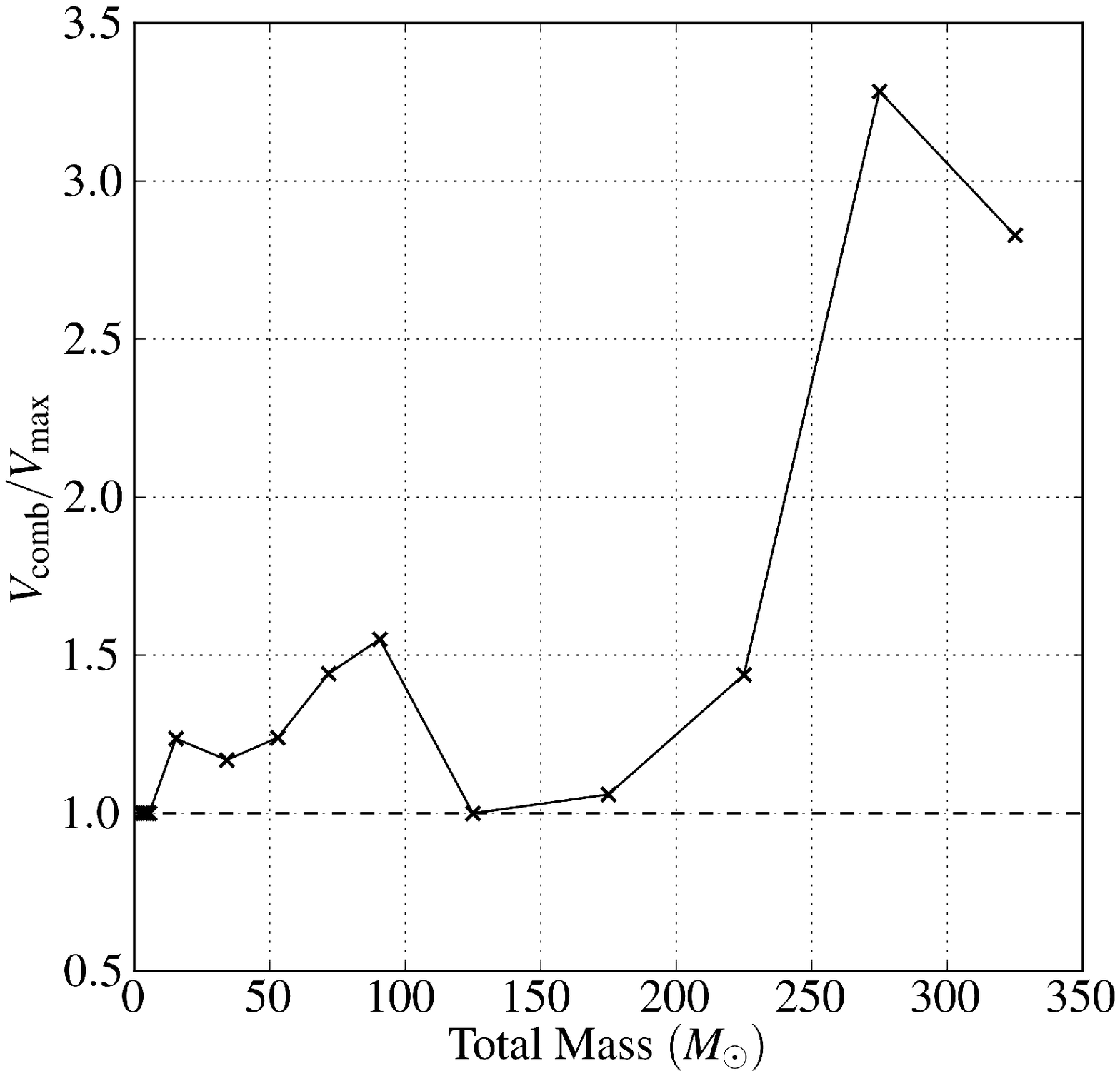}
\caption{The upper pane shows the cumulative 50\% efficiency contours at false-alarm rate of 0.28 events per year. The combined search contour envelops the single classifier contours from above.  The lower pane shows the plot of the ratio of the visible volume of the combined search to the visible volume of the most sensitive classifier in each mass bin. Having a ratio greater than one indicates that the combined search gains sensitivity across the entire mass range.}
\label{fig:efficiency_contours}
\end{figure}

For demonstration of these effects and further insight, we plot cumulative 50\% efficiency contours on the distance/total mass plane for signals to be detected above the threshold, Figure~\ref{fig:efficiency_contours}. The threshold is set by the lowest measured false-alarm rate of 0.28 events per year (corresponding to the left most point on the sensitivity curves in Figures~\ref{fig:bns_LVFAR}--\ref{fig:highmass_LVFAR}). In Figure~\ref{fig:efficiency_contours}, the contour for the combined search envelops contours of the other pipelines. Furthermore, we calculate the corresponding visible volume for the joined search and plot its ratio to the visible volume of the most sensitive pipeline in each mass bin, shown in the lower pane of Figure~\ref{fig:efficiency_contours}.

\section{Conclusions}
\label{sec:conclusions}

We consider the problem of combining outputs of partially redundant search methods analyzing the same data in the context of gravitational wave searches. We suggest that the likelihood ratio, Eq.~(\ref{eq:likelihoodratio}), provides a natural unified ranking for the candidate events identified by the search methods. It has a straightforward interpretation --- events
from each method are ranked according to the ratio of the method's sensitivity
to its background. After forming the joined list of candidate events, calculation of
the posterior probability distribution or an upper limit on the rate of
gravitational-wave emissions can proceed exactly as it would for a single search method. Therefore, there is never a problem consistently accounting for multiple trials in the analysis. If the combined search is interpreted as a
counting experiment, the procedure for calculating the upper limit from multiple searches is similar to that suggested in \cite{Sutton:2010}. In that case, classifiers contribute in proportion to their sensitivities. Additionally, our method accounts for information about the classifier's background, which is important when dealing with experimental data containing non-Gaussian noise artifacts. 

We test our procedure by simulating a search for gravitational waves from compact binary coalescence in the data from LIGO's S4 science run. We combine outputs from four search methods
--- the low-mass CBC, high-mass CBC, ringdown, and Coherent WaveBurst analysis pipelines ---
analyzing data with injected gravitational-wave signals from compact coalescing
binaries in a wide range of masses. We find that our algorithm is robust
against nuisance pipelines --- those that are not sensitive to the targeted
gravitational-wave sources. Moreover, the combined search proves to
have greater or comparable sensitivity to any individual pipeline. In the simulations, we observe that the pipelines we use contribute different events to the total
count of detected signals, thus increasing robustness and the overall probability of
detecting gravitational waves from coalescing binaries. This effect is
especially pronounced for sources in near to mid range distances. Overall, our simulations show that searches for gravitational waves from coalescing binaries can benefit from combining results of multiple analysis methods by means of the likelihood-ratio statistic.

\acknowledgments

Authors would like to thank Ilya Mandel and Jolien Creighton for many fruitful discussions and helpful suggestions. This work has been supported by NSF grants PHY-0600953 and PHY-0923409. DK was supported from the Max Planck Gesellschaft. LP and RV was supported by LIGO laboratory. JB was supported by the Spanish MICINN FPA2010-16495 grant and the Conselleria d’Economia Hisenda i Innovacio of the Govern de les Illes Balears. LIGO was constructed by the California Institute of Technology and Massachusetts Institute of Technology with funding from the National Science Foundation and operates under cooperative agreement PHY-0757058.

\appendix

\section{Multivariate likelihood-ratio statistic}
\label{appendix:JointLR}

The maximum likelihood-ratio procedure described in Section \ref{sec:method} does not account for potentially useful information contained in the correlations between classifiers. In this section, we derive the formal expression for the multivariate statistic that is optimal by construction and includes all available information. We discuss some of its properties and its relation to the maximum likelihood-ratio statistic, Eq.~(\ref{maxLRapprox}).

In the absence of internal thresholds, each classifier assigns a rank, $\data_i$, to every data sample. In this case, the vector of ranks, $\vec{\data} \equiv (\data_1, \data_2, \ldots \data_n)$, can be interpreted as the reduced experimental data, and the problem of combining searches is analogous to a detection problem. The general problem of detection in the presence of arbitrary noise was discussed in detail in \cite{UWMLSC:2010a}. Here, we state
the final result and refer the interested reader to \cite{UWMLSC:2010a} for
derivation and further discussion. The optimal solution, assuming the Neyman-Pearson criteria that requires the maximization of the signal detection probability at a fixed rate of false alarms, ranks data samples by
the likelihood-ratio detection statistic. For $n$ classifiers, this takes the form
\begin{equation}
\label{LRGenFormula}
\Lambda(\data_1,\data_2,\ldots,\data_n) = \frac{\int \! p(\data_1, \data_2,\ldots,\data_n \given \sig, 1)p(\sig \given 1)\diff \sig}{p(\data_1, \data_2,\ldots,\data_n \given 0)}\,,
\end{equation}
where $\sig$ stands for a gravitational-wave signal, $p(\data_1,
\data_2,\ldots,\data_n \given \sig, 1)$ is the probability distribution for the
vector of detection statistics $ (\data_1, \data_2,\ldots,\data_n)$ in the case
when the gravitational-wave signal $\sig$ is present in the data, $p(\data_1,
\data_2,\ldots,\data_n \given 0)$ is the analogous distribution for the noise,
and $p(\sig \given 1)$ is the distribution of signal parameters for the targeted
population of gravitational-wave sources.

The joint likelihood ratio, Eq.~(\ref{LRGenFormula}), includes the output
from all classifiers and by construction
provides the optimal ranking. We can simplify the expression in
Eq.~(\ref{LRGenFormula}) by noting that for $n=2$,
\begin{equation}
\begin{split}
\Lambda(\data_1,\data_2) =& \frac{\int \! p(\data_1, \data_2 \given \sig, 1)
p(\sig \given 1)\diff \sig}{p(\data_1, \data_2 \given 0)}\\
 =& \frac{1}{2}\Biggl(\frac{\int \! p(\data_1  \given \data_2,\sig, 1)
 p(\sig \given 1)\Lambda(\data_2, \sig)\diff \sig}{p(\data_1  \given \data_2,0)}\\
 & \phantom{\frac{1}{2}\Biggl(}+
 \frac{\int \! p(\data_2  \given \data_1,\sig, 1)
 p(\sig \given 1)\Lambda(\data_1, \sig)\diff \sig}{p(\data_2  \given \data_1,0)}\Biggr)\,,
 \end{split}
 \label{twopipelines}
\end{equation}
\noindent and extending linearly for $n>2$.

In practice, computing the conditional probabilities in
Eq.~(\ref{twopipelines}) can be a nontrivial task. Moreover, as the number of
classifiers increases, computing the necessary
conditional probabilities becomes a less and less viable option and it is
necessary to develop an approximation.  The maximum likelihood-ratio procedure can be regarded as such even though it does not follow from the multivariate expression (\ref{LRGenFormula}) in a straightforward way. In order to find a relation between the two, it is useful to
consider the limiting cases of Eq.~(\ref{twopipelines}).  First, assume there
is no correlation between the measurements made by each classifier. In that case, $p(\data_i \given \data_j,\sig,1) = p(\data_i \given \sig, 1)$ (and similarly for the denominators), so that
\begin{equation}
\label{uncorrelated}
 \Lambda(\data_1,\data_2) = \int \! \Lambda(\data_1, \sig)\Lambda(\data_2, \sig)p(\sig \given 1)\diff \sig \approx \Lambda(\data_1)\Lambda(\data_2) \,.
 \end{equation}

Factorization in the last step is justified because only one of the classifiers exhibits a non-trivial response in the presence of a signal in the data. As a consequence, the un-marginalized likelihood ratio, $\Lambda(\data, \sig)$, for that classifier is a function of the signal, $\sig$, sharply peaked around the true parameters of the signal, whereas the other likelihood ratio is almost constant.

\noindent At the other extreme, consider the case of two strongly
correlated classifiers. Then
\begin{equation}
\begin{split}
\frac{\int \! p(\data_1, \data_2 \given \sig)
p(\sig \given 1)\diff \sig}{p(\data_1, \data_2 \given 0)}
&= \frac{\delta(\data_1 - \data_2) \int \! p(\data_2 \given \sig)
p(\sig \given 1)\diff \sig}{\delta(\data_1 - \data_2)p(\data_2 \given 0)} \\
& =\Lambda(\data_1) \,.
\end{split}
\label{correlated}
\end{equation}
\noindent Both cases can be easily generalized for $n>2$ classifiers. 

When classifiers are strongly correlated, the maximum likelihood ratio is
trivially equivalent to (\ref{correlated}). In the opposite case, the absence
of correlation between the classifiers implies their complementarity. If one
classifier identifies a significant event, the others do not. This means that
typically only one of the likelihood ratios in the product in
Eq.~(\ref{uncorrelated}) will be significantly different from unity. Therefore,
in this case, picking the maximum of the single classifier likelihood ratios or
calculating their product has similar effect. Based only on these extreme
situations, it is difficult to determine how good of an approximation the
maximum likelihood ratio is in the intermediate case. Nevertheless, we
conjecture that the truly useful information can only be in correlations
between the classifiers using incomplete, but complementary, information about
the signal (e.g. template-based searches using inspiral and merger or ringdown
waveforms). Even in this situation, the inclusion of correlations should be a
next-order effect.

\section{Maximum likelihood-ratio statistic}
\label{appendix:maxLRapprox}

One can gain further insight into the statistic defined by Eq.~(\ref{maxLRapprox}) by mapping the ranks, $\data_i$, to their likelihood ratios, $\Lambda_i(\data_i)$. The mapping is defined by  Eq.~(\ref{eq:likelihoodratio}).  The data space of the combined search is  $\vec{\Lambda} \equiv (\Lambda_1, \Lambda_2, \ldots, \Lambda_N)$. For the $i^\mathrm{th}$ classifier, the probabilities of detection and of false alarm are given by
\begin{align}
P_1^i &= \int \! \Theta \left(\Lambda_i - \Lambda_i^*\right) p(\vec{\Lambda} \given 1) p(1) \diff \vec{\Lambda} \label{single_pipeline_P1} \\
P_0^i &= \int \! \Theta \left(\Lambda_i - \Lambda_i^*\right) p(\vec{\Lambda} \given 0) p(0) \diff \vec{\Lambda} \label{single_pipeline_P0} \,.
\end{align}
and for the combined search by
\begin{align}
\tilde{P}_1 &= \int \! \Theta \left(\max(\vec{\Lambda}) - \Lambda_\mathrm{c}^*\right) p(\vec{\Lambda} \given 1) p(1) \diff \vec{\Lambda} \label{combined_pipeline_P1} \\
\tilde{P}_0 &= \int \! \Theta \left(\max(\vec{\Lambda})  - \Lambda_\mathrm{c}^*\right) p(\vec{\Lambda} \given 0) p(0) \diff \vec{\Lambda} \label{combined_pipeline_P0} \,,
\end{align}
where $\Lambda_i^*$ and $\Lambda_\mathrm{c}^*$ are detection thresholds
determined from the threshold value for the false alarm probability, $P_0^*$,
which is the same for all classifiers.

\begin{figure}
\includegraphics[width=3in]{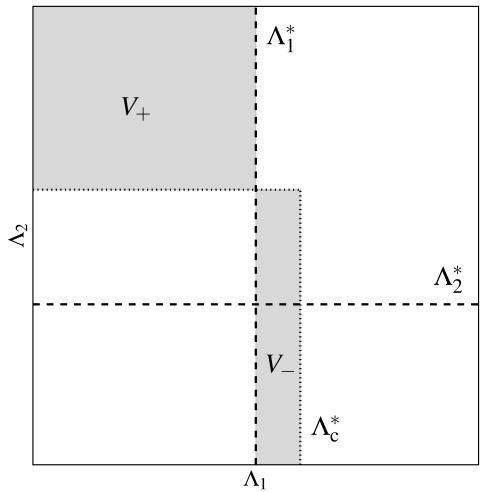}
\caption{The detection thresholds can be visualized for individual and combined searches in the $(\Lambda_1, \Lambda_2)$ space. For individual searches, the detection threshold appears as the vertical dashed line ($\Lambda_1^*$) and horizontal dashed line ($\Lambda_2^*$). The $\Lambda_\mathrm{c}^*$ threshold is the dotted line. The shaded regions represent the data points gained, $V_+$, and lost, $V_-$, by the combined search when referencing a search performed with $\Lambda_1$.}
   \label{fig:L1_vs_L2_diagram}
\end{figure}

The efficiency of the combined search is expected to be, at the very least, no less than the efficiency of any of the classifiers being used ($\tilde{P}_1 \geq P_1^i$). This is a necessary condition for the maximum likelihood-ratio procedure to be applicable. To get a better understanding of what this condition implies and when it is expected to hold, consider the simple case of combining a pair of classifiers. This can be generalized in a straightforward way to arbitrary number. The data space, in this case, is a positive quarter in the $(\Lambda_1, \Lambda_2)$ space. The lines of constant likelihood ratio, $\Lambda_i$, are horizontal or vertical lines. The lines of constant joint likelihood ratio, given by Eq.~(\ref{LRGenFormula}), can be complicated curves even in the $(\Lambda_1, \Lambda_2)$ plane and define the optimal detection surfaces. The corresponding surfaces for the maximum likelihood-ratio statistic form a square, centered at the origin, with sides parallel to the $\Lambda_1$ and $\Lambda_2$ axes. This configuration is visualized on Figure \ref{fig:L1_vs_L2_diagram}, where $\Lambda_1^*$ (vertical dashed line), $\Lambda_2^*$ (horizontal dashed line), and $\Lambda_\mathrm{c}^*$ (dotted line) are the thresholds corresponding to a particular value of the probability of false alarm, for single and combined searches respectively. Detection regions for each classifier consist of all points for which the argument of the theta function in the expressions for detection and false alarm  probabilities, Eqs. (\ref{single_pipeline_P1}) and (\ref{single_pipeline_P0}),   is positive. The detection region for the $i^{\mathrm{th}}$ classifier is defined by the condition $\Lambda_i > \Lambda_i^*$. All data points in the plane satisfying this condition are counted as detection of a signal. The detection region for the combined search defined by Eqs. (\ref{combined_pipeline_P1}) and (\ref{combined_pipeline_P0})  consists of the points satisfying two conditions: $\Lambda_1 > \Lambda_\mathrm{c}^*$ and  $\Lambda_2 > \Lambda_\mathrm{c}^*$. Recall that the false alarm probability for all searches is the same, which implies that
\begin{equation}
\label{FAP_condition}
\int_{V_i} \! p(\Lambda_1, \Lambda_2 \given 0 ) \diff \vec{\Lambda} = \int_{V_\mathrm{c}} \! p(\Lambda_1, \Lambda_2 \given 0) \diff \vec{\Lambda} \,,
\end{equation}
where $V_i$ and $V_\mathrm{c}$ denote the detection regions for either of the individual searches and for the combined
search respectively. This implies that $\Lambda_\mathrm{c}^*  > \Lambda_i$
---  the threshold for the combined search is higher than the threshold for any
of the individual pipelines. Indeed, if it was not true, then Eq.~(\ref{FAP_condition}) could not be satisfied since $V_i \subset V_\mathrm{c}$.  
This would correspond to moving the vertical dashed line to the right of the solid square in Figure
\ref{fig:L1_vs_L2_diagram}, as an example for classifier $\Lambda_1$.
Thus, the diagram shown in Figure~\ref{fig:L1_vs_L2_diagram} represents the only allowed
configuration. Continuing with the classifier $\Lambda_1$, one can identify the
set of points gained by the combined search, $V_+$, which are points not included
in $V_1$, and the set of points lost, $V_{-}$, those belonging to $V_1$ but not to
$V_\mathrm{c}$. Both sets are shown in Figure~\ref{fig:L1_vs_L2_diagram} as shaded
regions. It is clear that the efficiency of the combined search will be
greater or equal to the efficiency of the classifier $\Lambda_1$ if and only if
\begin{equation}
\label{DEP_condition}
\int_{V_+} \! p(\Lambda_1, \Lambda_2 \given 1) \diff \vec{\Lambda} \geq \int_{V_-} \! p(\Lambda_1, \Lambda_2 \given 1) \diff \vec{\Lambda} \,.
\end{equation}
Note that at the same time
\begin{equation}
\int_{V_+} \! p(\Lambda_1, \Lambda_2 \given 0) \diff \vec{\Lambda} = \int_{V_-} \! p(\Lambda_1, \Lambda_2 \given 0) \diff \vec{\Lambda} \,,
\end{equation}
by virtue of Eq.~(\ref{FAP_condition}). Thus, Eq.~(\ref{DEP_condition})
 states that the joint likelihood for the points in $V_+$
must be (on average) greater than or equal to the joint likelihood for the points in $V_-$. For this case, exchanging $V_+$ for $V_-$ results in a positive gain. It is not
unjustified to expect that Eq.~(\ref{DEP_condition}) would be satisfied
in most practical situations. After all, according to the classifier $\Lambda_2$, 
points in $V_+$ have a better chance of being a signal than those in $V_-$, because
$\Lambda_2$ for any point in $V_+$ is greater than $\Lambda_1$ for any point in
$V_-$. In effect, when combining searches using the maximum likelihood-ratio
approximation, one exchanges points from $V_-$ with decent $\Lambda_1$ and low
$\Lambda_2$ in favor of points in $V_+$ with low $\Lambda_1$ but high
$\Lambda_2$. Consider an extreme case when the classifier $\Lambda_2$ is not
informative. The probability of getting a high likelihood ratio in the
absence of the signal is very low. Then, the total probability $\int_{V_+} p(\Lambda_1,
\Lambda_2 \given 0)$ is negligible, effectively making $V_-$ a null set.
This implies robustness of the approximation against nuisance, non-informative
classifiers. The above steps can be mirrored for the classifier $\Lambda_2$, resulting in the same conclusions.

 In conclusion, we should stress that, although condition~(\ref{DEP_condition}) does not hold in general, it is expected to be satisfied when combining well designed classifiers that are sufficiently different to be able to complement each other's detection efficiencies. These are the typical cases that arise in practice.

\bibliography{../../bibtex/iulpapers}

\end{document}